\documentclass[twocolumn]{aastex631}
\usepackage{graphicx}
\usepackage{wasysym}
\usepackage{threeparttable}
\usepackage{booktabs}    
\usepackage{lipsum}      
\received{2024 June 7}
\revised{2024 July 18}
\accepted{2024 August 2}
\begin{document}

\title{Redshifted Sodium Transient near Exoplanet Transit}

\author[0000-0002-1655-0715]{Apurva V. Oza}
\affiliation{Division of Geological and Planetary Sciences, California Institute of Technology, Pasadena, USA}
\thanks{Contact e-mail: \href{mailto:oza@caltech.edu}{oza@caltech.edu}}
\affiliation{Jet Propulsion Laboratory, California Institute of Technology, Pasadena, USA}
\author[0000-0002-7990-9596]{Julia V. Seidel}
\affiliation{European Southern Observatory, Santiago de Chile, Chile}
\author[0000-0001-8981-6759]{H. Jens Hoeijmakers}
\affiliation{Department of Astronomy and Theoretical Physics, Lund Observatory, Lund, Sweden}
\author[0000-0001-6093-5455]{Athira Unni} 
\affiliation{Department of Physics and Astronomy, University of California, Irvine, Irvine, CA, USA} 
\affiliation{Indian Institute of Astrophysics, Bangalore, India}
\author[0000-0002-3239-5989]{Aurora Y. Kesseli}
\affiliation{Caltech/IPAC-NASA Exoplanet Science Institute, Pasadena, USA}
\author[0000-0002-6917-3458]{Carl A. Schmidt} 
\affiliation{Center for Space Physics, Boston University, Boston, USA}
\author[0000-0003-0891-8994]{Sivarani Thirupathi}
\affiliation{Indian Institute of Astrophysics, Bangalore, India}
\author[0000-0003-3355-1223]{Aaron Bello-Arufe}
\affiliation{Jet Propulsion Laboratory, California Institute of Technology, Pasadena, USA}
\author[0000-0002-0206-8231]{Andrea Gebek}
\affiliation{Sterrenkundig Observatorium, Universiteit Gent,  Ghent, Belgium}
\author[0000-0003-0446-688X]{Moritz Meyer zu Westram}
\affiliation{Physikalisches Institut, Universit\"{a}t Bern, Bern, Switzerland}
\author[0000-0001-9047-2965]{Sérgio G. Sousa}
\affiliation{Instituto de Astrofísica e Ciências do Espaço, Universidade do
Porto, CAUP,  Porto, Portugal} 
\author[0000-0002-7928-3167]{Rosaly M.C. Lopes}
\affiliation{Jet Propulsion Laboratory, California Institute of Technology, Pasadena, USA}
\author[0000-0003-2215-8485]{Renyu Hu}
\affiliation{Jet Propulsion Laboratory, California Institute of Technology, Pasadena, USA}
\affiliation{Division of Geological and Planetary Sciences, California Institute of Technology, Pasadena, USA}
\author[0000-0002-9068-3428]{Katherine de Kleer}
\affiliation{Division of Geological and Planetary Sciences, California Institute of Technology, Pasadena, USA}
\author[0000-0003-0652-2902]{Chloe Fisher}
\affiliation{Department of Physics, University of Oxford, Oxford, UK}
\author[0000-0002-7442-491X]{Sébastien Charnoz}
\affiliation{Universit\'{e} de Paris, Institut de Physique du Globe de Paris, CNRS, Paris , France}
\author[0000-0002-6525-7013]{Ashley D. Baker}
\affiliation{Department of Astronomy, California Institute of Technology, Pasadena, USA}
\author[0000-0003-1312-9391]{Samuel P. Halverson}
\affiliation{Jet Propulsion Laboratory, California Institute of Technology, Pasadena, USA}
\author[0000-0001-6720-5519]{Nicholas M. Schneider}
\affiliation{Laboratory for Atmospheric and Space Physics, University of Colorado Boulder, Boulder, CO, USA}
\author[0000-0002-4797-2419]{Angelica Psaridi}
\affiliation{Observatoire de Gen\`{e}ve, Universit\'{e} de Gen\`{e}ve, Gen\`{e}ve, Switzerland}
\author[0000-0001-9003-7699]{Aurélien Wyttenbach}
\affiliation{Observatoire de Gen\`{e}ve, Universit\'{e} de Gen\`{e}ve, Gen\`{e}ve, Switzerland}
\author[0000-0002-3150-8988]{Santiago Torres}
\affiliation{Institute of Science and Technology Austria (ISTA),  Klosterneuburg, Austria} 
\author[0009-0006-7723-0519]{Ishita Bhatnagar}
\affiliation{Birla Institute of Technology and Science, Pilani, India}
\author[0000-0001-7798-5918]{Robert E. Johnson}
\affiliation{Department of Physics, New York University, USA}
\affiliation{University of Virginia, Charlottesville, USA}

\begin{abstract}
Neutral sodium (Na I) is an alkali metal with a favorable absorption cross section such that tenuous gases are easily illuminated at select transiting exoplanet systems. We examine both the time-averaged and time-series alkali spectral flux individually, over 4 nights at a hot Saturn system on a $\sim$ 2.8 day orbit about a Sun-like star WASP-49 A. Very Large Telescope/ESPRESSO observations are analyzed, providing new constraints. We recover the previously confirmed residual sodium flux uniquely when averaged, whereas night-to-night Na I varies by more than an order of magnitude. On HARPS/3.6-m Epoch II, we report a Doppler redshift at $v_{ \Gamma, \mathrm{NaD}} =$ +9.7 $\pm$ 1.6 km/s with respect to the planet's rest frame. Upon examining the lightcurves, we confirm night-to-night variability, on the order of $\sim$ 1-4\% in NaD rarely coinciding with exoplanet transit, not readily explained by stellar activity, starspots, tellurics, or the interstellar medium.  Coincident with the $\sim$+10 km/s Doppler redshift, we detect a transient sodium absorption event dF$_{\mathrm{NaD}}$/F$_{\star}$ = 3.6 $\pm$ 1 \% at a relative difference of $\Delta F_{\mathrm{NaD}} (t) \sim$ 4.4 $\pm$ 1 \%,  enduring $\Delta t_{\mathrm{NaD}} \gtrsim$ 40 minutes. Since exoplanetary alkali signatures are blueshifted due to the natural vector of radiation pressure, estimated here at roughly $\sim$ -5.7 km/s, the radial velocity is rather at +15.4 km/s, far larger than any known exoplanet system. Given that the redshift magnitude v$_{\Gamma}$ is in between the Roche limit and dynamically stable satellite orbits, the transient sodium may be a putative indication of a natural satellite orbiting WASP-49 A b. 
\end{abstract}

\section{Introduction} \label{intro_wasp49}
Sodium and potassium are alkali metals, observed in the vapor form, which are remarkably strong absorbers in transmission spectroscopy \citep{seagersasselov2000}. Neutral sodium gas (Na I) at a transiting exoplanet system for instance, may be probed at tenuous densities with line-of-sight columns far below N $\ll$ 10$^{12}$ Na/cm$^{-2}$ \citep{draine2011}, roughly 0.01 picobars or $\sim$4 nanopascals for an isothermal atmosphere on an Earth-sized planet. The favorable cross-section of Na I enabled its first detection at the hot Saturn HD209458 b \citep{charbonneau02} in low-resolution by \textit{Hubble Space Telescope}. New high-resolution precision by Very Large Telescope (VLT)/ESPRESSO since identified previously interpreted planetary sodium as stellar from HD209458 \citep{Casasayas-Barris2021}. It was suggested shortly after the initial exoplanetary detections, that since neutral sodium  is photoionized within minutes \citep{huebnermujherjee} on a few days orbit, compared to the $\sim$ few hours transit duration of transiting exoplanets \citep{johnson06b}, alkali metals may be exogenic. In other words, since sodium is uniquely probed as a neutral, a continuous supply of exogenic neutral atoms may be preferred, especially at high altitudes above the planet's surface, where Na I is probed. 
 
Since sodium and potassium are D-line resonance doublets, the D$_2$/D$_1$ ratio also allows for spectral characterization \citep{go2020}, which so far has helped constrain a range of possible sodium densities systematically at up to 10 exoplanet systems where neutral sodium has been reliably identified \citep{Langeveld_2022}.  Here we examine one of the dozens of known alkali exoplanet systems (or transiting planetary systems with reported alkali detections) WASP-49 b, reported to host a high-altitude layer of neutral sodium at roughly $\sim$ 1.5 R$_p$, thought to be thermal and non-isothermal \citep{wyttenbach17, cubillos_2017, oza2019b, fisher_heng2019, go2020}. Large-aperture, low-resolution observations similar to the successful \textit{James Webb Space Telescope} (JWST) also allow for coarse analyses to assess the presence of alkalis. We note for instance observations with the VLT/FORS2 \citep{Lendl2016}, and recently the Gran Telescopia Canarias (GTC/OSIRIS) \citep{Jiang2023} do not detect previously identified Na I or K I, motivating the need for several more transit observations, and in-depth high-resolution analyses.

\par WASP-49 A (2MASS 06042146–1657550) is a faint (m$_G$ = 11.29 $\pm$ 0.1 ) Sun-like star discovered by the Wide Angle Search for Planets (WASP) \citep{Pollacco2006_WASP} now known to be in a binary star system common to most transiting exoplanets \citep{Mugrauer2019}. Subsequently in 2012, its transiting Saturn-mass companion WASP-49 A b was discovered \citep{lendl2012} and characterized by VLT/FORS2 \citep{Lendl2016}. 
    The hot Saturn exoplanet system transits its Sun-like star every 2.78 days or 66.7 hours for a duration of 2.15$^{+0.02}_{-0.01}$ hours or $\sim$ 128 minutes. This transit timescale with a precision of roughly 1 minute, over exposure times of $\sim$ 10 minutes defines our temporal uncertainties to describe neutral sodium as we will describe below. Here, we seek to probe the Hill sphere of WASP-49 A b in NaD light with three nights of HARPS/ESO 3.6-m data and new VLT/ESPRESSO 8.2-m data. The Hill sphere is the gravitational sphere of influence where natural satellites are expected to be stable for $\sim$ Gyrs within $\sim \frac{1}{2}$ Hill sphere, as estimated by several dynamical works \citep{domingos06, cassidy09, Kisare2023}. This is below an altitude of $\sim$ 2.0 R$_{p}$ for WASP-49 A b, and above the Roche limit for a rocky body at $\sim$ 1.2 R$_{p}$ \citep{oza2019b}. 
    Na I at the Jupiter system is ultimately sourced by volcanic NaCl from its satellite Io \citep{kuppers2000} sputtering into a ``banana cloud" of order $\sim$ R$_J$ \citep{schneiderbrown81, haff81}. A fast component largely escapes Jupiter $\gtrsim$ 100 km/s, forming a $\sim$1000 R$_J$ sodium cloud \citep{mendillo90}. Jupiter's gravitationally bound component $\lesssim$ 60 km/s is fueled by two plasma processes: (1) ``jet" localized to Io from dissociative recombination of NaCl$^{+}$ formed via charge-exchange with photoionized gas in Io's ionosphere \citep{Schmidt_2023}, and (2) ``stream" extending beyond $\gtrsim$10 R$_J$ neutralized by ions Na$^{+}$ or NaX$^{+}$ entrained in the diffuse Io plasma torus far from Io \citep{wilson2002}.
 Our high-resolution approach, led by the initial detection at HD189733b \citep{Redfield_2008} is able to identify Doppler-shifted alkali absorption, in time. Since Doppler shifts in Jovian Na I spectra pinpoint the radial velocity of Io and its ion chemistry \citep{Cremonese1992}, we are in principle, able to indirectly trace natural satellites. 
New star-planet system parameters probed by our new high-resolution VLT/ESPRESSO observations  are reported in Table \ref{TablePARAMETERS}, while Table \ref{TableOBSERVATIONS} indicates the start of each exoplanet transit epoch. Based on our new large-aperture observations of the star, we constrain a smaller stellar mass at M$_{\star}$= 0.894 $\pm$ 0.04 M$_{\odot}$ and a slightly smaller planet mass M$_{p}$= 0.365 $\pm$ 0.019 M$_{J}$ than previous studies \citep{Stassun2017AJ, wyttenbach17, lendl2012}.  We first present our transmission spectra observations in Section \ref{section:2}, upon analysis we confirm neutral sodium at the resonance line doublet in Section \ref{sec:spectra} (time-averaged) and Section \ref{sec:lightcurves} (sodium lightcurves). Finally, we discuss possible interpretations in Section \ref{sec:discussion}.

\section{Neutral Alkali Metal Observations} \label{section:2}
We present 4 nights of time-resolved high-resolution data for WASP-49 A near exoplanet transit. The first three epochs of observations began on UT 2015 December 6, 2015 December 31, and 2016 January 14 using HARPS/ESO 3.6-m \citep{Mayor2003, harps2006SPIE}, and on 2020-12-16 using VLT/ESPRESSO \citep{Pepe2021}. In \S \ref{sec:spectra}, we present time-averaged transmission spectra results centered at the sodium D resonance lines (Na D at $\lambda_{D_2}$= 5889.950 and $\lambda_{D_1}$=5895.924 Å), and in \S \ref{sec:lightcurves}, we construct lightcurves of neutral sodium.

\begin{table*}[ht]
    \centering
    \caption{Star-planet system parameters. RV designates radial velocity. Barycentric velocity corrections are provided as ranges for each individual epoch (I $\rightarrow$ IV) in km/s.}
    \resizebox{\textwidth}{!}{%
    \begin{threeparttable}
    \begin{tabular}{@{}llr@{}}
    \toprule
    \textbf{Parameter} & \textbf{Variable} & \textbf{Value} \\
    \midrule
    \multicolumn{3}{l}{\textit{Stellar Parameters}} \\
    Name & WASP-49 A & -- \\
    \textit{Gaia} DR3 ID$^{a}$ & GDR3ID & 2991284369063612928 \\
    Apparent Magnitude$^{a}$ [mag] & $m_G$ (\textit{Gaia}) & $11.29 \pm 0.10$ \\ 
    Spectral Type & G6V & -- \\
    Stellar Age$^{\ddagger}$ [Gyr] & $t_{\star}$ & $7 \pm 2$ \\
    Stellar Photosphere Temperature$^{\ddagger}$ [K] & $T_\mathrm{eff}$ & $5543 \pm 61$ \\
    Stellar Gravity (spec)$^{\ddagger}$ [cm~s$^{-2}$] & $\log g$ & $4.37 \pm 0.10$ \\
    Stellar Gravity (Gaia)$^{\ddagger}$ [cm~s$^{-2}$] & $\log g$ & $4.38 \pm 0.03$ \\
    Stellar Metallicity$^{\ddagger}$ [dex] & $[\mathrm{Fe/H}]$ & $-0.08 \pm 0.04$ \\
    Stellar Micro Velocity$^{\ddagger}$ [km~s$^{-1}$] & $v_\mathrm{mic}$ & $0.79 \pm 0.02$ \\
    Stellar Macro Velocity$^{\ddagger}$ [km~s$^{-1}$] & $v_\mathrm{mac}$ & $2.91 \pm 0.09$ \\
    Stellar Radial Velocity$^{\ddagger}$ [km~s$^{-1}$] & $vsini$ & $1.93 \pm 0.25$ \\
    Stellar Mass$^{\ddagger}$ [$M_\odot$] & $M_{\star}$ & $0.894^{+0.039}_{-0.035}$ \\
    Stellar Radius$^{b}$ [R$_{\odot}$] & R$_{\star}$ & $0.976 \pm 0.034$ \\
    Stellar Density$^{c}$ [g/cm$^3$] & $\rho_{\star}$ & $1.43 \pm 0.19$ \\
    Stellar Rotational Velocity$^{d}$ (K$_1$) & v$_{\star,rot}$ & $57.5 \pm 2.1$ m/s \\
    \midrule
    \multicolumn{3}{l}{\textit{Planet Parameters}} \\
    Exoplanet Transit Midpoint$^{e}$ [BJD] & T$_{0}$ & $2457377.596934 \pm 0.000080$ \\
    Planet/Star Radius Ratio$^{e}$ & R$_p$/R$_{\star}$ & $0.116 \pm 0.0007$ \\ 
    Impact Parameter$^{d}$ & b & $0.7704^{+0.0072}_{-0.0077}$ \\ 
    Planet Transit Duration$^{\ddagger}$ [hours] & t$_{14}$ & $2.146^{+0.022}_{-0.007}$ \\ 
    Planet Mass$^{\ddagger}$ [M$_J$] & M$_p$ & $0.365 \pm 0.019$ \\ 
    Planet Radius$^{f}$ [R$_J$] & R$_p$ & $1.115 \pm 0.047$ \\ 
    Planet Orbital Period$^{e}$ [days] & $\tau_P$ & $2.78173691 \pm 0.00000014$ \\
    Planet Density$^{d}$ [g/cm$^3$] & $\rho_p$ & $0.288 \pm 0.006$ \\
    Radiative Equilibrium Temperature [K] & T$_{p,eq}$ & 1400 \\
    \midrule
    \multicolumn{3}{l}{\textit{Velocity Parameters}} \\
    Planet Orbit (RV semi-amplitude) [km/s] & K$_p$ & 151.6 \\
    System (star+planet) RV [km/s] & v$_{\gamma}$ & $41.7261 \pm 0.0011$ \\
    Barycentric velocity Epoch I [km/s] & $v_{bary,I}$ & $6.20 \rightarrow 6.92$ \\
    Barycentric velocity Epoch II [km/s] & $v_{bary,II}$ & $-3.87 \rightarrow -3.07$ \\
    Barycentric velocity Epoch III [km/s] & $v_{bary,III}$ & $-9.3 \rightarrow -8.6$ \\
    Barycentric velocity Epoch IV [km/s] & $v_{bary,IV}$ & $2.65 \rightarrow 3.05$ \\
    \bottomrule
    \end{tabular}
    \begin{tablenotes}
        \footnotesize
        \item $^{\ddagger}$ This work. \quad $^{a}$ \citet{GAIADR3} \quad $^{b}$ \citet{lendl2012} \quad $^{c}$ \citet{Stassun2017AJ} \quad $^{d}$ \citet{Wyttenbach2017} \quad $^{e}$ \citet{Kokori2023} \quad $^{f}$ \citet{Bonomo2017}
    \end{tablenotes}
    \end{threeparttable}
    }
    \label{TablePARAMETERS}
\end{table*}

Our ESPRESSO/VLT 8.2-m observations provide a first large-aperture epoch compared to three HARPS/3.6-m epochs monitoring the exoplanet Hill sphere (see Table \ref{TableOBSERVATIONS}). A second ESPRESSO/VLT epoch was disturbed by strong winds during planetary ingress, however the in-transit and egress data remain useful in evaluating the stellar continuum and activity, as well as measuring fundamental star-planet properties, such as the system metallicity $\chi_{\mathrm{Fe/H}}$ following \citet{Sousa2008} (Table \ref{TablePARAMETERS}). We use only 1 VLT/ESPRESSO epoch for the transmission spectrum.
In the first VLT/ESPRESSO transit (2020-12-15), the first four exposures with shorter exposure times (300s instead of 360s) had approximately 50\% lower flux and were discarded for consistency. Additionally, three epochs of HARPS data, previously analyzed by \citet{wyttenbach17}, were re-analyzed similar to the ESPRESSO data. The signal-to-noise ratio range, SNR56, is provided for the 56th échelle order near the NaD doublet at 5889.95 and 5895.92 Å. To correct for telluric features around the sodium doublet, we employed {\tt molecfit}, version 1.5.1 \citep{Smette2015, Kausch2015}, demonstrated to remove Earth's water vapor lines near the sodium doublet in several works, e.g. HD189733b \citep{Allart2017}, WASP-76b \citep{seidel2019}, WASP-121b \citep{hoeijmakers2020}, WASP-166 \citep{Seidel2020b}, WASP-127, \citep{Seidel2020c}, HAT-P-57b, KELT-7b, KELT-17b, KELT-21b, MASCARA-1b, and WASP-189b \citep{Stangret2022}. 
We combined all spectra in the observer's rest frame and verified that the telluric lines were corrected down to the noise level of the combined spectrum. No telluric sodium emission or Moon contamination was detected in fiber B (on-sky observations). Moreover, we carefully inspected Earth's (telluric) rest frame velocities (Table \ref{TablePARAMETERS}, $v_{bary}$) for Na I and did not detect any, ruling out sodium at Earth's barycenter.
The data were analyzed following the procedure at WASP-76 b \citep{Seidel19} in the NaD wavelength range. Significant scatter was observed due to flux differences at the planetary sodium position relative to stellar sodium lines. Stellar sodium absorption reduces the signal-to-noise ratio if planetary sodium overlaps with the same wavelength range. ESPRESSO exposures during transit had shorter exposure times (360s compared to 600s for HARPS), causing flux to be dominated by red noise at the center of stellar sodium line cores. 
The average spectral flux was 1500, while the flux in the stellar line core was 30-120, a decrease by one to two orders of magnitude. Similar studies masked the trace of the stellar sodium line \citep{Barnes2016, Borsa2018, Seidel2020b, Seidel2020c}; we also masked a window of 4 km/s in the stellar rest frame across all four epochs for consistency.
The Rossiter-Mclaughlin (RM) effect, generally a concern when evaluating Na I \citep{Nuria2020}, describes the additional shift in wavelength for transmission spectra due to the rotation of the star as the planet crosses the stellar surface. 
\citet{wyttenbach17} investigated various instrumental, telluric, and stellar influences on the transmission spectrum and used the reloaded RMmethod \citep{cegla16} to estimate $<$ 100 ppm or $\lesssim$0.01 \% of a RM effect, concluding either a pole-on transit or indeed an extremely slow-rotating host star, with $v sin i_{\star} \lesssim$ 0.46 km/s.  
Therefore, the RM effect did not have a measurable impact on the HARPS transmission spectrum as it was below the noise level for ESPRESSO data.
Upon reduction, we reproduced the time-averaged transmission spectra and detection levels from both \citet{wyttenbach17} and \citet{Langeveld2022}. Notably, time-averaged alkali depths may vary based on flux-averaged bin widths, a factor carefully examined by \citet{Langeveld2022} for 10 alkali exoplanet systems, alongside variable D$_2$/D$_1$ alkali ratios from optically-thick \citep{seagersasselov2000} to optically-thin systems \citep{go2020}. We note that while our time-averaged signature over the three epochs (\S \ref{sec:spectra}) agrees largely with \citet{wyttenbach17, Langeveld2022}, our time-resolved signature does not (\S \ref{sec:lightcurves}). This is likely due to our choice of not discarding exposures we believe were significant SNR when examining their propagated error in the lightcurves (\S \ref{sec:lightcurves}). 

\begin{table*}
    \centering
	\begin{tabular}{lccccccr} 
\hline
     \textbf{EPOCH} & \textbf{UT Date} & \textbf{Instrument} & \textbf{$\#$Spectra$^{1}$} & \textbf{$\Delta$ t$_{\mathrm{exp}}$ [minutes]} & \textbf{SNR56} & \textbf{Airmass$^{2}$} & \textbf{Seeing}  \\
      \hline
      I & 2015-12-06 & HARPS$^{\dagger}$ & 41(14/27) & 10 & $21-33$ & $1.9-1.0-1.3$ & not recorded  \\
      
      II & 2015-12-31 & HARPS$^{\dagger}$ & 42(12/30)$^3$ & 10 & $10-38$ & $1.4-1.0-2.2$ & $0.5-1.3$  \\
      III & 2016-01-14 & HARPS$^{\dagger}$ & 40(13/27) & 10 & $22-40$ & $1.2-1.0-2.3$ & $0.5-1.0$  \\
      IV & 2020-12-16 & ESPRESSO$^{\ddagger}$ & 62(36/26) & 6 & $21-26$ & $1.60-1.15-1.00$ & $0.6$   \\
      V & 2021-02-18$^4$ & ESPRESSO$^{\ddagger}$ & 17(12/5) & 8.3 & $32-27$ & $1.15-1.65$ & $0.5-1.0$  \\
      \hline
    \hline
	\end{tabular}
	\caption{\textbf{ High-resolution observations.} 
   $^1$ \# Spectra parentheses show the total (in- /and  out-of-) transit spectra after processing exposures.
 SNR56 range and $^2$ airmass at the beginning, centre, and end of each transit.
 $^3$ Two out-of-transit exposures were rejected due to emission features either stemming from cosmic rays or the instrument.
 $^4$ Partial transit + egress. Planetary ingress was not observed due to wind, rendering the baseline too short for meaningful analysis. The duration of each epoch is indicated in hours. } 
 {$^\dagger$  \texttt{ESO program ID 096.C-0331; PI: Ehrenreich.}$^{\ddagger}$ \texttt{ESO Program ID Prog.ID = 106.21QX.001, 106.21QX.003 ; PI:Oza}.} \\
 	\label{TableOBSERVATIONS}
\end{table*}
The high-resolution WASP-49 A b spectrum \textit{averaged} over three epochs is shown in several independent works \citep{wyttenbach17, cubillos17, fisher_heng2019, oza2019b, Langeveld2022}, however, here we seek to examine each night \textit{individually}. 
We separate our analysis technique across two components centered at the sodium doublet: \S \ref{sec:Doppler2} NaD spectra (time-averaged sodium)  \& \S \ref{sec:lightcurves} NaD lightcurves (instantaneous sodium). 
\begin{figure*}[ht]
    \centering
    \includegraphics[width=\textwidth]{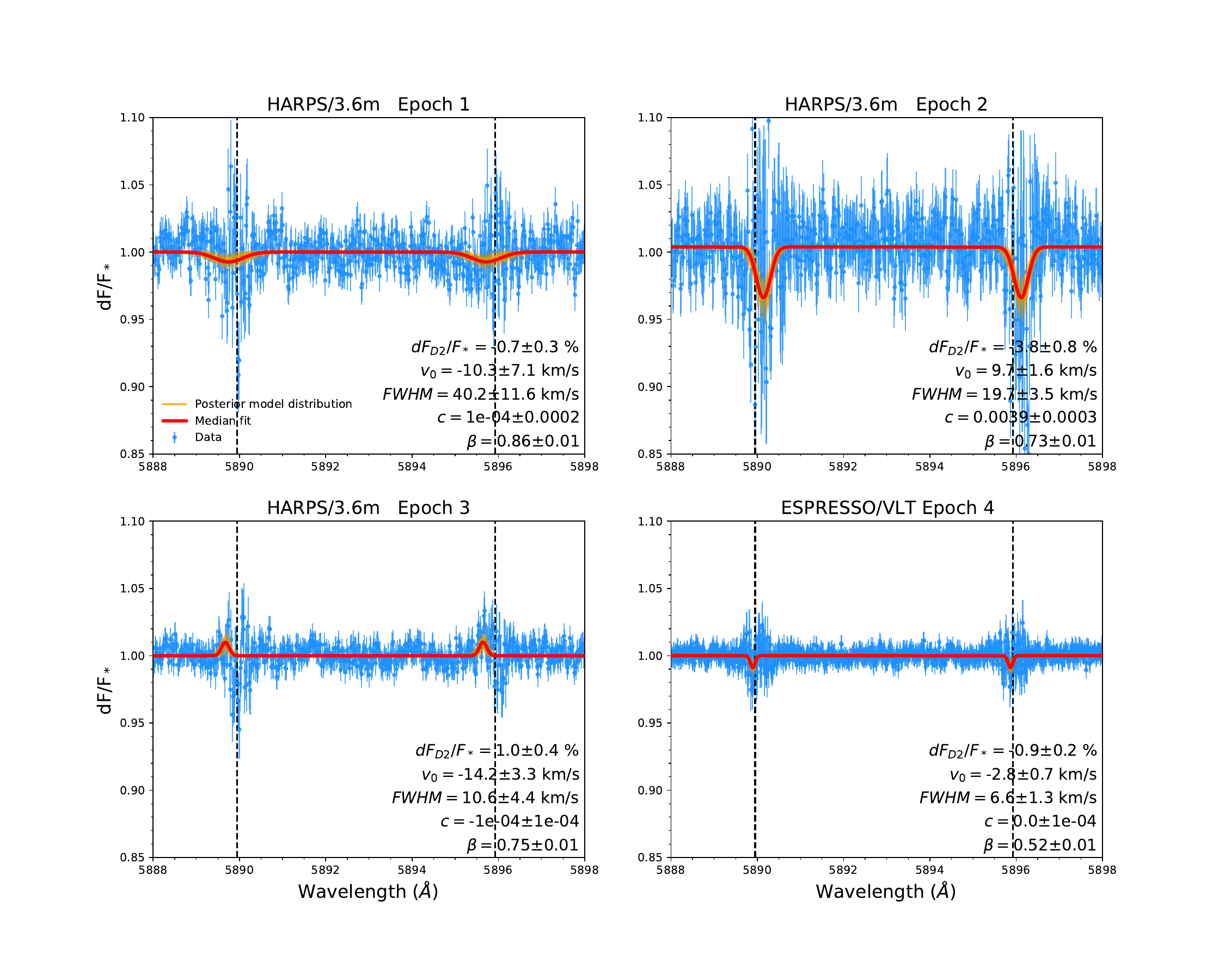}
    \caption{\textbf{Transmission spectrum of neutral atomic sodium (Na I) over four epochs.} Raw data (blue), posterior model (orange), and median fit (red).  (a) Epoch I: 2015 December 6; (b) Epoch II: 2015 December 31; (c) Epoch III: 2016 January 14; (d) Epoch IV: 2020 December 15 (ESPRESSO/VLT). The Epoch II \textit{net} Doppler shift is measured to be a redshift v$_0$ = +9.7 km/s $\pm$ 1.6 km/s. Epoch III is a non-detection. Epoch I shows a blueshift at low significance. Companion residual lightcurves for each epoch are provided (Fig. \ref{fig:LC}).}
    \label{fig:SPECTRUM_multipanel}
\end{figure*}

\section{Time-Averaged Sodium}\label{sec:spectra}
\begin{figure*}
    \centering
    \begin{minipage}{0.44\textwidth}
        \centering
        \includegraphics[width=\linewidth]{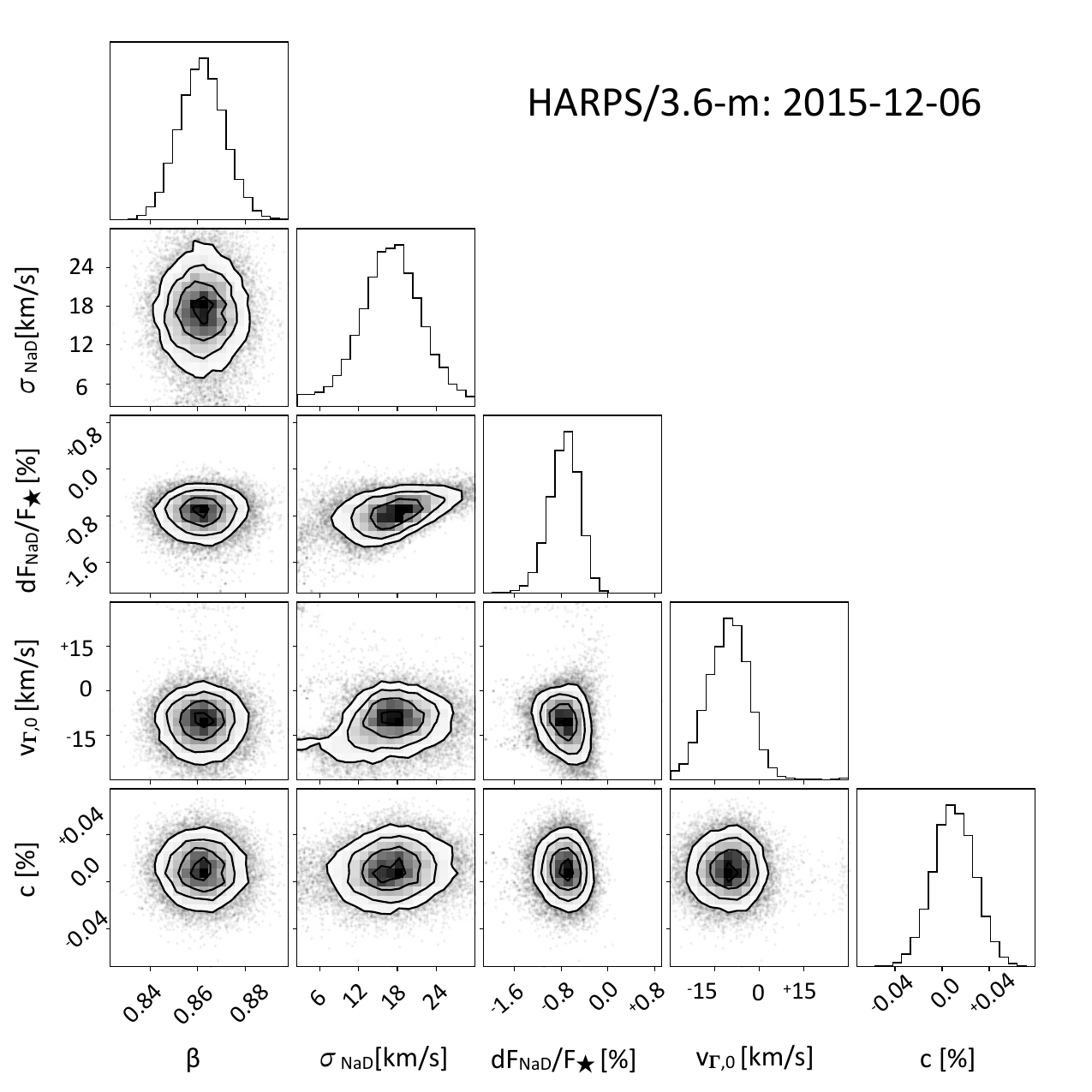}
    \end{minipage}
    \hfill
    \begin{minipage}{0.44\textwidth}
        \centering
        \includegraphics[width=\linewidth]{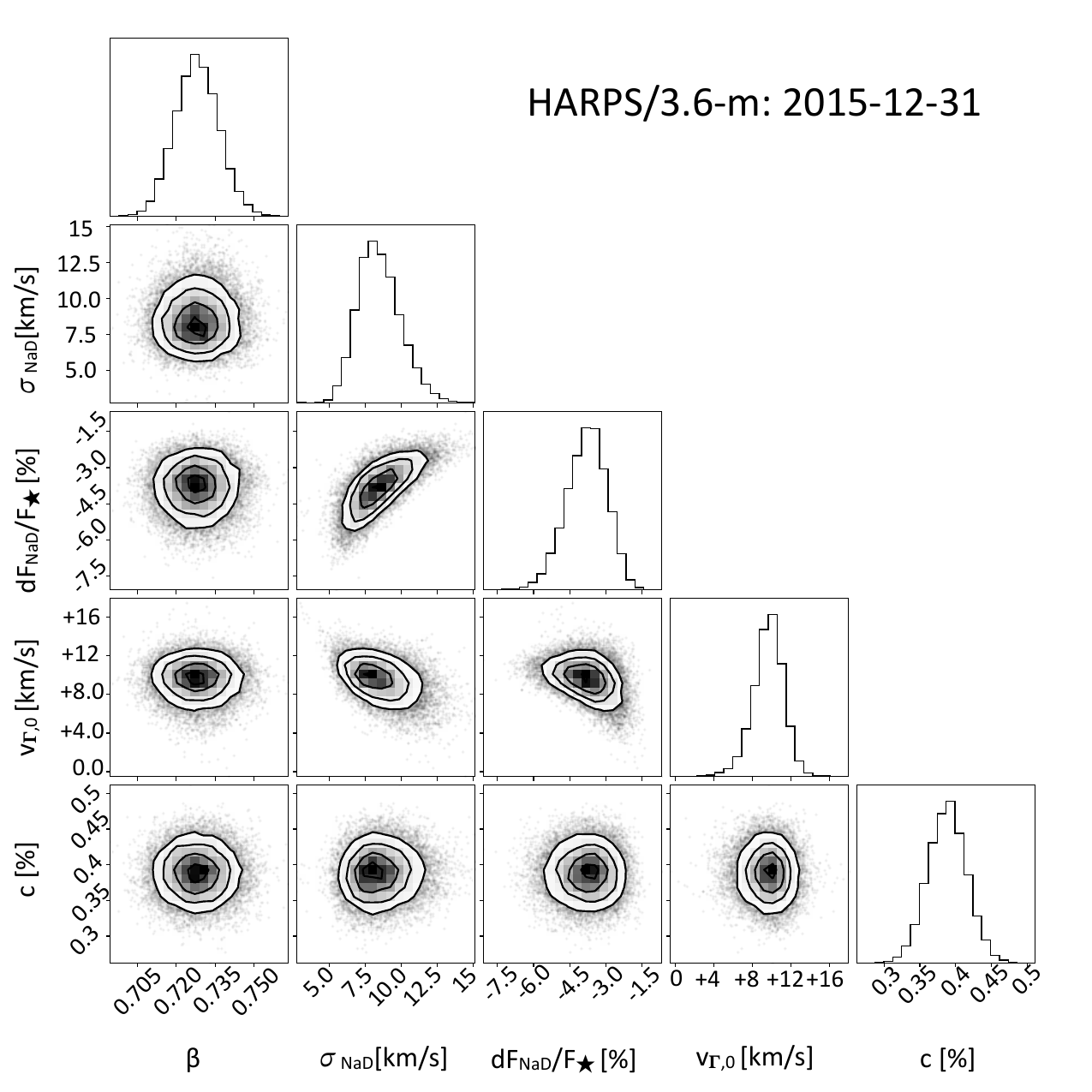}
    \end{minipage}
    \vspace{0.5cm}
    \begin{minipage}{0.44\textwidth}
        \centering
        \includegraphics[width=\linewidth]{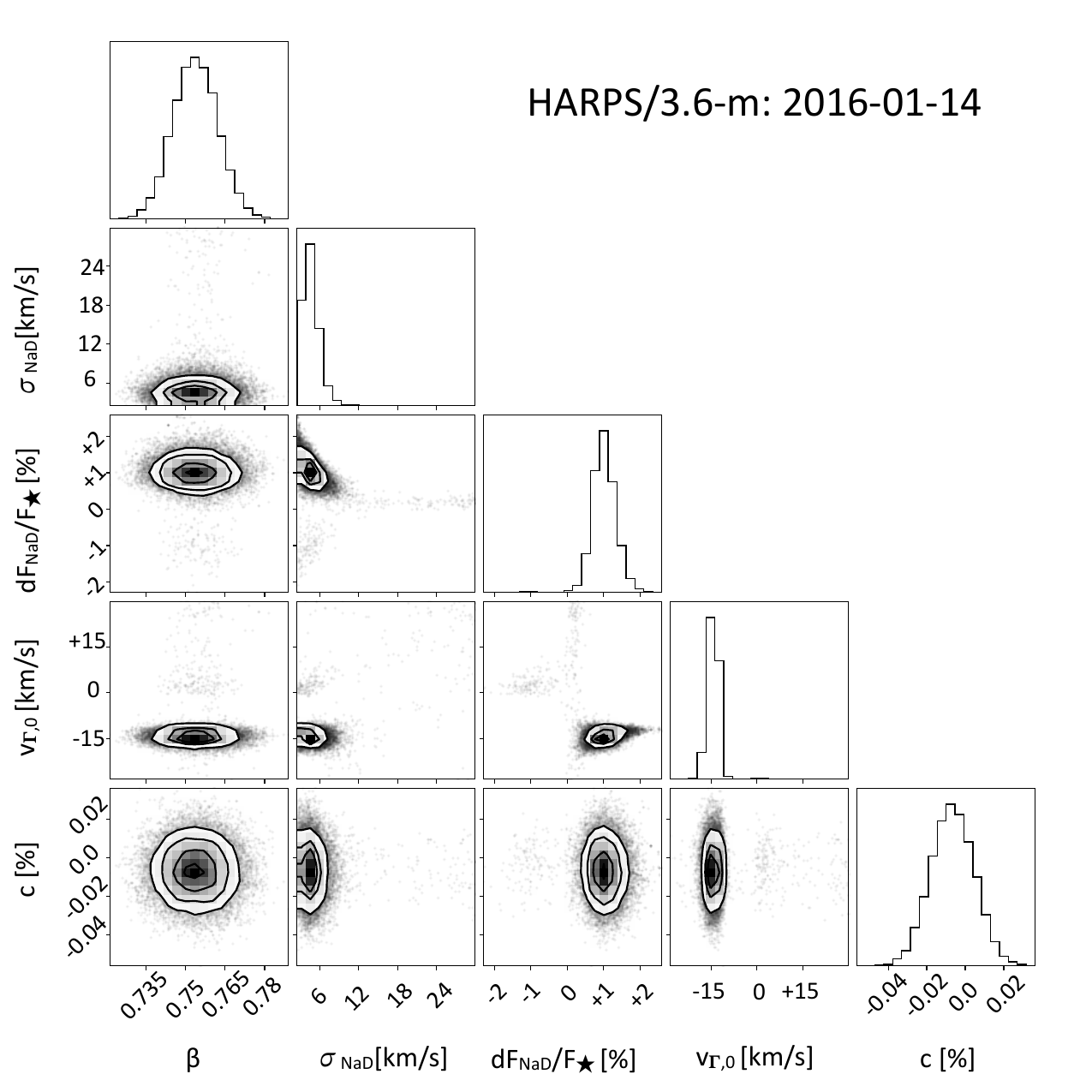}
    \end{minipage}
    \hfill
    \begin{minipage}{0.44\textwidth}
        \centering
        \includegraphics[width=\linewidth]{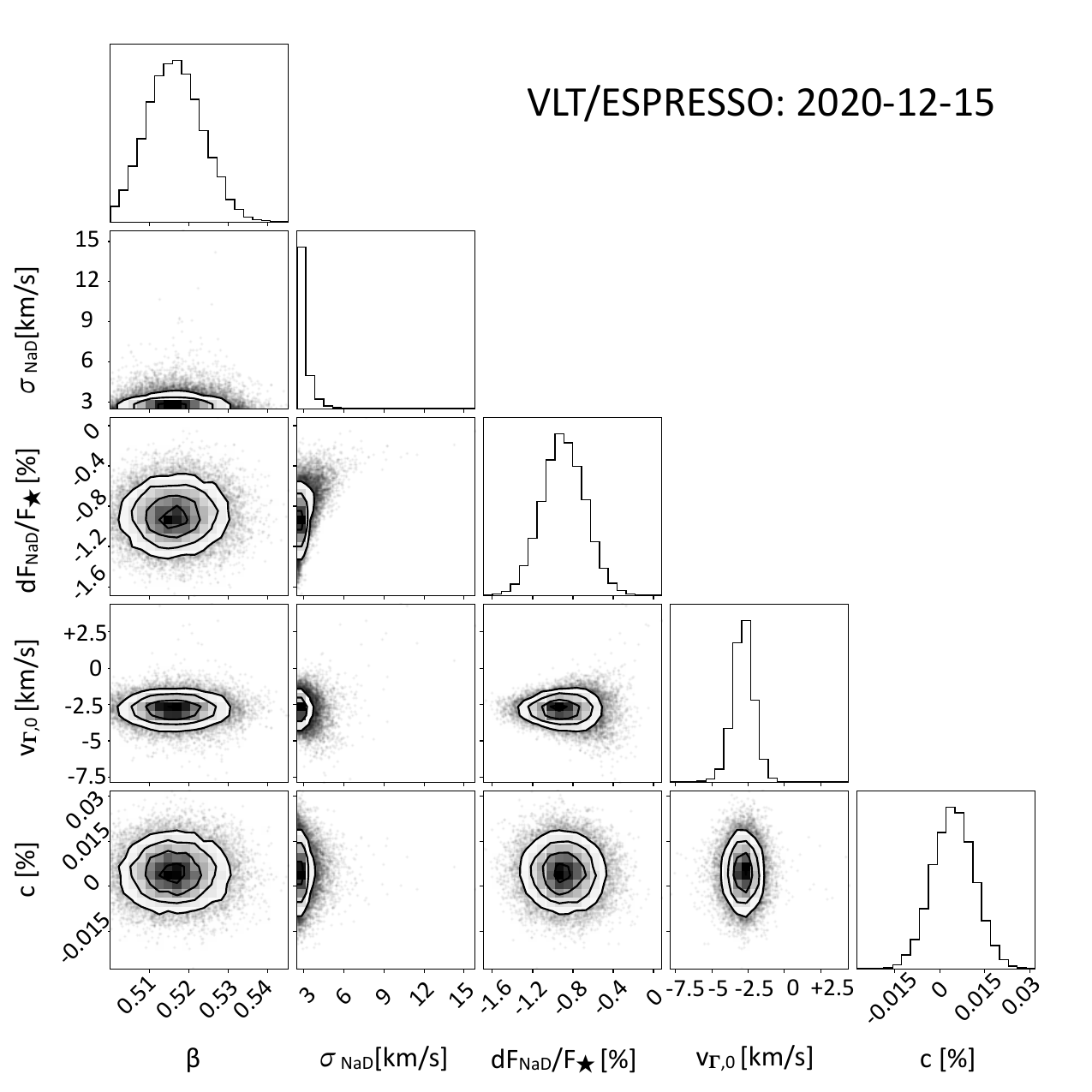}
    \end{minipage}
    \caption{Four-panel Gaussian fit runs for a nominal sodium absorption solution where $f_{\mathrm{D_2/D_1}}$= 1.0. (a) Epoch I: 2015 December 6 (b) Epoch II: 2015 December 31 (c) Epoch III: 2016 January 14. (d) Epoch IV: 2020-12-15. Posterior distributions of the model parameters describing the Na doublet as a pair of Gaussians with the same amplitude $dF_{Na}/F_{\star}$, centroid velocity shift $v_0$, width $\sigma$ and continuum $c$. The uncertainties are scaled with $\beta$ which is a free fitting parameter.  }
    \label{fig:cornerplots}
\end{figure*}

We build four transmission spectra following \citep{Seidelwasp172, Seidel23, Allart2020}. Here, we focus on alkali line profiles and their apparent variability. We perform a Bayesian retrieval framework using a No U-Turn Sampler \citep{Betancourt2017} implemented with \texttt{Jax} and \texttt{NumPyro}  \citep{jax2018github,bingham2019pyro,phan2019composable}. To model the doublet, we assume that both lines are Gaussian

with centroid velocity $v_0$ and width $\sigma$, with a flat continuum at $c$,  and the dimensionless parameter $\beta$ is used to scale the uncertainties, indicating that the errors have been marginally over-estimated. Overall, we see distinct Na I variability in the sodium flux absorption amplitudes from epoch to epoch: d$\mathcal{F}_{Na}/\mathcal{F}_{\star}$ (Epoch I) = -0.8 $\pm$ 0.3 \% ,  d$\mathcal{F}_{Na}/\mathcal{F}_{\star}$ (Epoch II) =  -3.8 $\pm$ 0.8 \% , d$\mathcal{F}_{Na}/\mathcal{F}_{\star}$ (Epoch III) =  +0.4 $\pm$ 0.5 \%, d$\mathcal{F}_{Na}/\mathcal{F}_{\star}$ (Epoch IV) = -0.8 $\pm$ 0.2 \%.  Best-fit models are shown in Figure \ref{fig:SPECTRUM_multipanel} along with posterior distributions in Figure \ref{fig:cornerplots}. 
The sodium flux on HARPS Epoch II on 2015 December 31, detected at roughly 5-$\sigma$, appears to be the primary source of the time-averaged sodium flux \citep{wyttenbach17, Langeveld2022}.  HARPS Epoch III on 2016 January 14 is consistent with a null-hypothesis (zero flux absorption at the sodium doublet) with a p$_{dF(\lambda)}$-value = 0.22. On the other hand, HARPS Epoch I \& II, 
and ESPRESSO Epoch IV have a non-trivial p$_{dF(\lambda)}$-value = 10$^{-2.4}$, 10$^{-5.9}$, and 10$^{-4.5}$ respectively. 
Based on the average D$_2$ line absorption, we compute the minimum neutral line-of-sight sodium column density based on the identified flux absorption at NaD$_2$ line center \citep{draine2011} are: N$_{\mathrm{Na}, 1} (\lambda) \sim$ 10$^{10.5 \pm 0.4}$ cm$^{-2}$ , N$_{\mathrm{Na}, 2} (\lambda) = $ 10$^{11.2 \pm 0.7}$ cm$^{-2}$ , N$_{\mathrm{Na}, 3} (\lambda)<$ 10$^{10.2 \pm 0.1}$ cm$^{-2}$ , and N$_{\mathrm{Na}, 4} (\lambda) \sim $ 10$^{10.5 \pm 0.4}$ cm$^{-2}$ .
Order-of-magnitude sodium variability is not expected for stable planetary atmospheres. Given that Epoch III was at higher signal-to-noise it is surprising that HARPS did not identify a time-averaged flux signature based on the column density of N$_{\mathrm{Na}, 3}$. Since the distinct epoch to epoch variability is not well known for exoplanets, we investigate several hypothesis that may contribute an alternative sodium signature along our line-of-sight. Furthermore, as demonstrated by \citep{Langeveld2022} measureable differences in line depths from independent reductions \citep{wyttenbach17} can occur due to binning, which further motivates time-dependent investigations. A bandpass of 0.75 Å was used in the current analysis, whereas \citet{Wyttenbach2017} used 0.4 Å. Choices on discarding certain exposures may lead to the current indication of emission in Night 3. However, as we will see in \S \ref{sec:lightcurves}, examining the data in time, is likely more useful for validating independent exposures. 
\begin{table*}[ht]
    \centering
    \caption{Posterior values from Figure \ref{fig:SPECTRUM_multipanel} are tabulated here with 1-$\sigma$ Gaussian uncertainties. dF$_{Na, \lambda}$/F$_{\star}$ (\%) is the absorption flux at sodium D-line center, v$_0$ is the Doppler shift from the exoplanet's rest frame,  $c$ is the continuum offset in parts-per-million, and $\beta$ is a normalization parameter, $\sigma_V$ is the FWHM of the line width.}
    \resizebox{\textwidth}{!}{%
    \begin{tabular}{lcccc}
    \toprule
    \textbf{Parameter} & \textbf{Epoch I} & \textbf{Epoch II} & \textbf{Epoch III} & \textbf{Epoch IV} \\ 
    \midrule
    dF$_{Na, \lambda}$/F$_{\star}$ (\%) & -0.736$\pm$0.263 & -3.768$\pm$0.833 & 1.006$\pm$0.355 & -0.896$\pm$0.212 \\
    v$_0$ (km/s) & -10.28$\pm$7.1 & +9.72$\pm$1.6 & -14.18$\pm$3.26 & -2.84$\pm$0.75 \\
    c (ppm) & 99.78$\pm$176.35 & 3886.68$\pm$271.43 & -67.28$\pm$112.73 & 45.97$\pm$69.44 \\
    $\beta$ & 0.86$\pm$0.01 & 0.73$\pm$0.01 & 0.75$\pm$0.01 & 0.52$\pm$0.01 \\
    $\sigma_V$ (km/s) & 17.09$\pm$4.92 & 8.35$\pm$1.5 & 4.52$\pm$1.88 & 2.79$\pm$0.56 \\
    \bottomrule
    \end{tabular}
    }
    \label{JensTable}
\end{table*}




\subsection{Spurious Stellar Sodium Signals}
Care is taken during our analysis to consider stellar NaD absorption, stellar lightcurve NaD absorption, and interstellar medium (ISM) absorption. The 11th magnitude Sun-like star system WASP-49 A (G6V) is not identified as an active-star unlike brighter transiting alkali exoplanet systems e.g. HD189733 (K-type star). Photometry from TESS and EulerCam show no significant flares, especially in NaD. Furthermore, unlike the brighter alkali exoplanet system HD209458b,  WASP-49 A b is minimally affected by the Rossiter McLaughlin effect \citep{Casasayas-Barris2018, Langeveld2022} with a trivial vsini $\sim$ 0.9 $\pm$ 0.3 km/s . Other systems such as WASP-189 b \citep{Prinoth2023} with larger vsin$i$ $\sim$ 93.1 $\pm$ 1.7 km/s \citep{Lendl2020}, and HAT-P-70 b \citep{Bello-Arufe_2023} at vsin$i$ $\sim$ 99.87 $\pm$ 0.65 km/s \citep{Zhou2019} also exhibit alkali sodium local to the planetary systems.  Regarding ISM sodium features, \citet{Langeveld2022} reported significant and narrow, ISM sodium absorption for several exoplanets, including WASP-21 and WASP-189, with intensities of $\gtrsim$90 \%. A Na I survey along 1005 lines-of-sight within 300 pc was extensively studied by \citep{Lallement2003, Welsh2010}. Of course, if the residual sodium absorption is indeed due to ISM, the signature would be identical for all transits, which is not the case c.f. \S \ref{sec:lightcurves}. Indeed, no such features were seen for WASP-49 A \citep{Langeveld2022}. Misaligned with the expected RV of the ISM relative to Earth's barycenter, a secondary sodium peak is Doppler blueshifted from the stellar rest frame at $\sim$ -19.5 km/s, matching the feature reported by \citet{wyttenbach17} at roughly -20 km/s. Unlike ISM-contaminated stars, this feature is weaker at only $\sim$ 10\%. Several more nights of long-baseline, large-aperture, échelle observations at the sodium doublet are needed to characterize the variability of this sodium feature, potentially linked to the identified Doppler shifts.

A stellar spot can be ruled out based on time-dependent stellar spectra from HARPS/3.6-m, VLT/ESPRESSO (2021-10-11), and UVES (2016-01-07) observations. The stellar sodium remains identical in each frame and wavelength, making a spurious event, or stellar spot crossing unlikely. Control experiments by \citet{wyttenbach17} using Mg I (5183.604 Å) and Ca I lines (6122.217 Å, 6162.173 Å) also rule out stellar chromospheric activity. WASP-49 A hosts a binary companion $\sim$ 0.34 M$_{\odot}$, separated by $\sim$ 443 AU from recent \textit{Gaia} analyses \citep{Mugrauer2019}. Far more observations are necessary to understand binary star dynamics' implications, however fiber-fed spectroscopy does indeed resolve WASP-49 A. Telluric sodium contamination is not detected in our spectral reduction, at Earth's reference frame. In the next section we look deeper into the astrophysical origin of the neutral sodium signature by examining radial velocity signatures. 


\subsection{Doppler Redshift of Neutral Sodium}\label{sec:Doppler2} 
We investigate the significance of Doppler shifts that appear to be present in some of the nights to learn more about the residual sodium signatures. Figure \ref{fig:cornerplots} in particular indicates a Doppler shift in Epoch II. At a larger significance than Epoch II's $\sim$ 5-$\sigma$ flux absorption (\S \ref{sec:spectra}), the significance of the radial velocity detection in Epoch II is larger at $\sim$ 6-$\sigma$. The sodium doublet is significantly red-shifted in Epoch II, at a radial velocity of $+9.7 \pm 1.6$ km/s. This is surprising as substantial alkali redshifts have not been measured at exoplanet systems, as most neutral atoms would be preferentially blueshifted $\sim$ -5 km/s for H$_2$O at HD189733 b \citep{Blain2024, Boucher2021, Brogi2016} and Na I \citep{Redfield_2008}. HD189733b also exhibits variable Doppler shift signatures in Na I \citet{loudenwheatley15} from v$_{\Gamma}$ = -5.3$^{+1.0}_{-1.4}$ $\rightarrow$ +2.3$^{+1.3}_{-1.5}$ km/s. 

In fact, high-resolution time-series data of Io's sodium in eclipse \citep{schmidt23}, is a solar system demonstration that substantial Doppler \textit{redshifts} of an alkali metal at a transiting system are uniquely fueled by natural satellites. Io fuels Jupiter's sodium exosphere out to a radius of $\sim$ 500 R$_p$ \citep{mendillo90}. Mercury's planetary sodium tail which extends to $\sim$ 1000 R$_p$ \citep{schmidt10} is again, only observed to be Doppler blueshifted and not redshifted due to radiation pressure. At the distance of WASP-49 A b, accounting for radiation pressure, the net system velocity of the residual sodium is at $+15.4 \pm 1.6$ km/s. This sodium velocity corresponds to a synchronous orbit at an altitude of $\sim$ 2.44 R$_p$ at roughly $\sim$ 19 hours.

If astrophysical, the null-hypothesis for Epoch II's Doppler shift (p$_{\Gamma}$-value = 10$^{-9.2}$) yields a less-than-a-billion chance for the Doppler redshift of optically thick Na I to be spurious. Similarly, if we force the Gaussian spectral fit to D$_2$/D$_1>$ 1 (in the optically thin regime) we still detect a significant Doppler redshift up to $+$12 km/s, with a preferred value near $+$6 km/s, although we do not believe this solution is physical.  For further credence regarding a redshifted radial velocity component, and more certainty regarding unknown systematics affecting the sodium absorption signature, we construct relative sodium lightcurves below in \S \ref{sec:lightcurves}. This is in supplement to previous analyses of residual sodium at this system, which did not time-resolve the Na D$_2$ line \citet{wyttenbach17, Langeveld2022}. By examining the time-evolution of sodium, we are able to pinpoint when in time (exposure times of 5-10 minutes) the observed redshift occurred relative to the exoplanet's transit across the stellar surface. 

\section{Sodium Light Curves}\label{sec:lightcurves} 
\begin{figure*}
\centering
  \centering
  \includegraphics[width=0.95\textwidth]{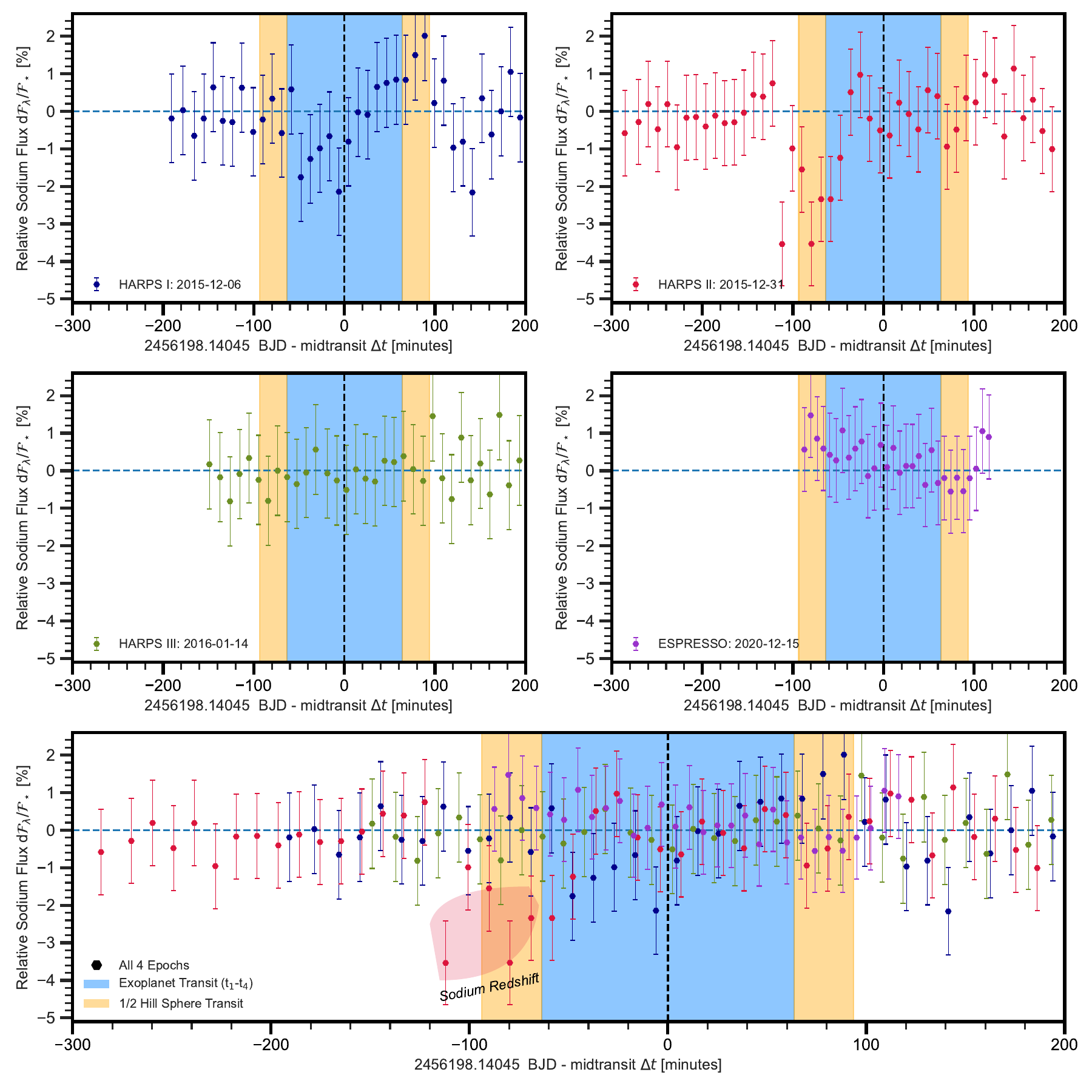} 
  \caption{\textbf{HARPS/ESPRESSO neutral sodium residual lightcurves (d$\lambda$=0.75 Å).} Black data points are Na D$_2$ spectra at four epochs centered at the transit of the hot Saturn WASP-49 A b, t$_{0, exoplanet}$. Yellow bands mark 1/2 of the Hill sphere transit time and blue bands mark the t$_1$-t$_4$ exoplanet transit time.   }   
  \label{fig:LC}
\end{figure*}

By separating the four observations in time, we are able to examine the evolution of the sodium gas, and directly probe \textit{when} the Doppler shifted sodium appears in terms of the known exoplanet midcenter time t$_0$. The ephemeris, accurate to sub-minute precision \citep{Kokori2023}, then provides tight constraints on when the exoplanet surface and its Hill sphere begin to transit. In this way, residual lightcurves may rule out symmetric atmospheric or magnetospheric geometries in time.
 
Relative lightcurves for four individual transits of WASP-49b are extracted \textit{in time} at 1-$\sigma$ standard deviation in Figure~\ref{fig:LC}, where the signal-to-noise (SNR) ratio is indicated in Table \ref{TableOBSERVATIONS}. 
The lightcurves are extracted by averaging the relative absorption signal of the Na D$_2$ and D$_1$ lines (over two bands with bandwidth $\Delta\lambda=0.75$\,\AA), and normalizing the transit depth to continuum bands without absorption lines \citep{wyttenbach17}. 

To not bias the feature seen in the second HARPS transit, HARPS relative absorption lightcurves
were generated with a master-out spectrum only taking into account spectra taken far away from
ingress and egress excluding each time $\delta$ = 0.015 in phase before and after the transit to exclude a sodium absorption source linked to the planet’s trajectory present in our master-out spectrum.

The four panels show that the relative sodium absorption signal is highly variable and transient, changing between epochs, as well as within the individual transits for the first two nights, and is not at all associated with the planetary transit window of $\sim$ 129 minutes, unlike other alkaline exoplanets. For instance, sodium at WASP-76b (also taken with HARPS/3.6-m) and WASP-172b (VLT/ESPRESSO) steadily absorbs at $\lesssim$0.5 \% relative to the much brighter star at higher SNR, and is more or less within the exoplanet transit window over three individual nights on HARPS \citep{Seidel19} and two nights of ESPRESSO \citep{Seidelwasp172}. 

In stark contrast, our current candidate WASP-49 A b does not have consistent absorption near t$_0$ in each transit. In fact the sodium lightcurve is nearly flat in certain epochs. 
The vanishing 2-4 \% signal in epochs three and four is difficult to reconcile with absorption from a stable planetary atmosphere or magnetosphere \citep{Lai2010_wasp12, Czesla2022}. The minimum change in the sodium column from Epoch II to Epoch III is: $\times$ 9.5. 

For Epoch II, we examine the SNR at 5500 Å to evaluate potential false positives. The 6-$\sigma$ Doppler redshift occurs close to $\sim$ 03:20 UT, long before exoplanet ingress at 05:22 UT near the sodium event, where we do identify a decreased SNR. Weather conditions in ESO ASM archives find no clouds detected on the pyrgeometer (``clear"), sufficient wind speed ($\sim$4 m/s), and stable temperatures ($\sim$15ºC) and humidity below 30\%.   We conclude that while the transit depth of the feature may be marginally overestimated, it is not an artifact due to atmospheric conditions (e.g. clouds). One possible hypothesis for an increase in seeing is a larger jet stream velocity along our line of sight which is known to affect seeing \citep{Vernin1986} and degrade SNR.

Moreover, in Epoch II the signal peaks midway through the ingress of $\sim 1/2$ the Hill sphere (orange bands in Figure~\ref{fig:LC}), long \textit{before} first contact of the planet (blue bands indicate first t$_1$ and fourth t$_4$ contact of the planet in Figure~\ref{fig:LC}). The first event occurs for a duration of 10 minutes, and the second event of sodium for another $\Delta t \sim$ 40 minutes, coincident with the exoplanet's Hill sphere and Roche limit transit time. The feature is indicated shaded in red in Figure \ref{fig:LC} When examining the radial velocity residuals at the sodium doublet in Epoch II, we indeed find that the large Doppler redshift is coincident with the $\gtrsim$ 40 minute, sodium feature occurring in the Hill sphere, over a minimum precision on the order of the exposure time $\sim$ 10 minutes.

\section{Conclusion} \label{sec:discussion}

We report significant kinematic and flux variability of neutral sodium absorption that does not coincide with the known exoplanet transit time of the hot Saturn WASP-49 A b.
The transient sodium detection on the night of UT 2015 December 31 (exoplanet transit midpoint at UT 2016 January 1 05:22) persists for over $\approx$ 40 minutes at a relative 4.4-$\sigma$ significance \textit{in temporal}, 4.8-$\sigma$ \textit{in spectral}, and 6.0-$\sigma$ in \textit{RV} space. While the 2015 December 6 feature is less significant at a relative flux of 2.7-$\sigma$, it too diminishes over $\approx$ 40 minutes prior to transit midcenter. The significances for other nights are low, primarily due to the dimness of the distant star at an apparent magnitude m$_v$ = 11.35.

Searches for companion ions and neutrals (e.g. S, O, Mg, Si) in optical and IR are encouraged to test the electrodynamics of rapidly ionized sodium \citep{Koskinen2014}. The presence of neutral potassium at the K I doublet at 7664.90 Å and 7698.96 Å was searched for with ESPRESSO however not detected, as expected, as it is far more tenuous than its companion alkali signature (Na I) in Epoch IV. The Keck Planet Finder, as well as future high-resolution instruments on the Thirty-Meter-Telescope, the Giant Magellan Telescope, or the Extremely Large Telescope, will help us pinpoint the origin of the transient metal signatures reported here.

If a variable sodium source exists within the planetary system, we cannot be sure of its orbit nor its cloud geometry based on 5 transits alone. More transits are needed and would be especially efficient since a single transit duration endures $\sim$ 2.14 hours + 1h of ingress \& egress, it monitors the entirety of the planetary Roche limit \citep{Roche_1873} and the $\sim \frac{1}{2}$ Hill sphere gravitational boundary discussed in \S \ref{intro_wasp49}. If satellite orbits endure only $\sim$ 6-11 hours, our observations probe up to $\sim$ 28-55\% of putative orbits per observation. A viable explanation may be that a satellite's sodium is uniquely illuminated at particular phases, for instance at maximum insolation during occultation or planetary shadow during eclipse. Circumstellar disk observations have revealed evaporating bodies too small to otherwise be seen, by exogenic Na I at LkH$\alpha$ 234 \citep{Chakraborty2004}, $\beta$ Pictoris \citep{Hobbs1985, Beust1990, Beust2024}, and possibly 55 Cancri-e \citep{Ridden-Harper2016, MeierValdes2023}. NaD absorption due to putative natural satellites and their accompanying clouds/tori have been suggested only by time-averaged spectra at several exoplanet systems thus far \citep{oza2019b, Gebek_2020, hoeijmakers2020, schmidt22, MeyerzuWestram2024}, however here we present an initial time-resolved signature. Both time-averaged and time-resolved cases encourage significant observational follow-up and study of tidal resonances, some of which is underway \citep{TokadjianPiro2023, Barr2023}. The gas variability is similar to ultraviolet, visible, or infrared observations of alkali, molecular oxygen, and water lines seen at Jupiter's satellites \citep{brown96, thomas96, retherford07, Thomas2022, dekleer2023}.

\section*{Acknowledgements}
The research described in this paper was carried out in part at the Jet Propulsion Laboratory, California Institute of Technology, under a contract with the National Aeronautics Space Administration. © 2024. California Institute of Technology. Government sponsorship acknowledged. 
AVO and JVS thank M. Lendl for constraints and discussions on the mass of WASP-49 A b. S.G.S acknowledges the support from
FCT through Investigador FCT contract nr. CEECIND/00826/2018 and POPH/FSE (EC).
\section*{Data Availability}
ESO data is available publicly on ESO's \href{http://archive.eso.org/eso/eso_archive_main.html}{science archive facility}. 

\bibliography{exomoons}{}
\bibliographystyle{aasjournal}

\end{document}